\pdfoutput=1

\documentclass[11pt]{article}

\usepackage[preprint]{acl}

\usepackage{bm}

\usepackage{CJKutf8}
\usepackage{makecell}
\usepackage{graphicx}
\usepackage{amsmath}
\usepackage{amssymb}
\usepackage{algorithm, algpseudocode}
\usepackage{hyperref}
\usepackage{multirow}
\usepackage{booktabs}
\algrenewcommand\algorithmicrequire{\textbf{Input:}}
\algrenewcommand\algorithmicensure{\textbf{Output:}}
\usepackage{multicol}
\usepackage{xcolor}
\usepackage{tcolorbox}

\usepackage{CJK} 

\definecolor{rq}{HTML}{1B365C}
\definecolor{rqBack}{HTML}{9ECBF7}

\usepackage{times}
\usepackage{latexsym}

\usepackage[T1]{fontenc}

\usepackage[utf8]{inputenc}

\usepackage{microtype}

\usepackage{inconsolata}

\usepackage{graphicx}

\title{Contextualized Token Discrimination for Speech Search Query Correction}

\author{Junyu Lu$^1$, Di Jiang$^1$, Mengze Hong$^2$, Victor Junqiu Wei$^3$, Qintian Guo$^4$, Zhiyang Su$^4$\\
$^1$AI Group, WeBank,
$^2$Hong Kong Polytechnic University\\
$^3$Macau University of Science and Technology\\
$^4$Hong Kong University of Science and Technology
}

\begin{document}
\maketitle

\begin{abstract}
Query spelling correction is an important function of modern search engines since it effectively helps users express their intentions clearly. With the growing popularity of speech search driven by Automated Speech Recognition (ASR) systems, this paper introduces a novel method named Contextualized Token Discrimination (CTD) to conduct effective speech query correction.
In CTD, we first employ BERT to generate token-level contextualized representations and then construct a composition layer to enhance semantic information.
Finally, we produce the correct query according to the aggregated token representation, correcting the incorrect tokens by comparing the original token representations and the contextualized representations.
Extensive experiments demonstrate the superior performance of our proposed method across all metrics, and we further present a new benchmark dataset with erroneous ASR transcriptions to offer comprehensive evaluations for audio query correction.
\end{abstract}

\section{Introduction}

Speech search has become a core feature in dialogue and interactive systems, like smartphones and intelligent vehicles, driven by user demand for seamless information access \cite{10.1145/2956235,ZhuNZDD21,torbati2021you}. These search engines process tens of millions of queries daily and rely heavily on automatic speech recognition (ASR) to convert spoken queries into transcripts for retrieving relevant documents or passages \cite{hafen2012speech, jiang2015teii}. However, ASR systems frequently generate transcripts with word errors \cite{tang2019yelling,song2019l2rs}. Fuzzy spelling and diverse accents can degrade performance, while character similarities often lead to misspellings. As a result, speech search operates with imperfect texts and queries~\cite{hafen2012speech}, potentially misrepresenting user intent and causing retrieval failures and user dissatisfaction.

\begin{figure}[!t]
\centering

\includegraphics[width=0.9\columnwidth]{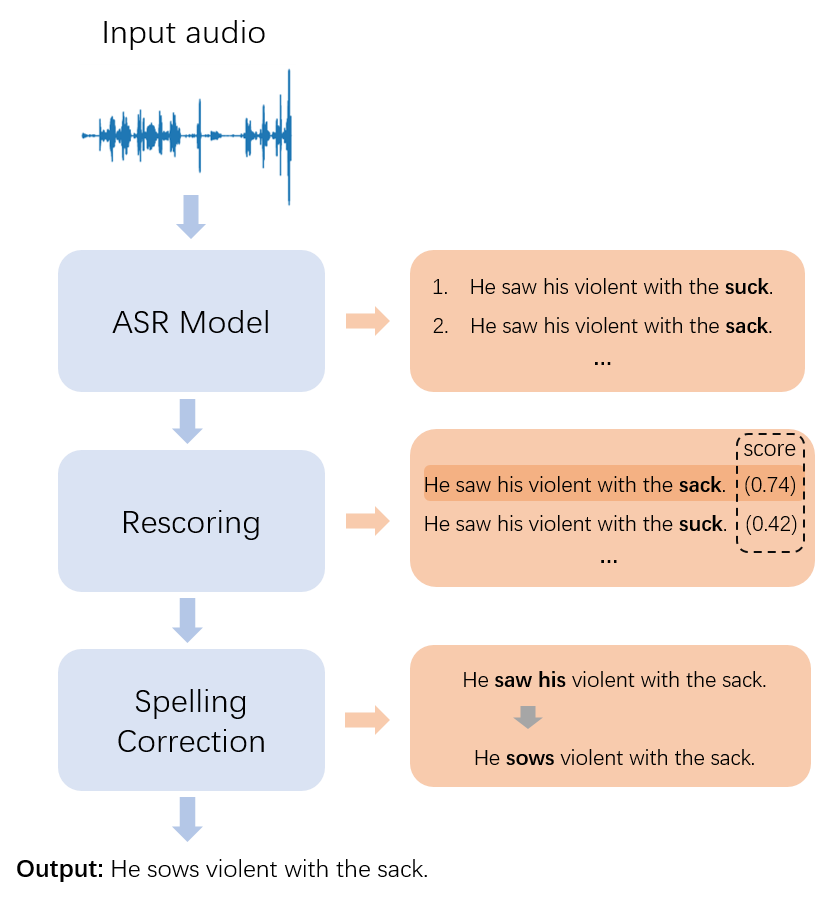}
\caption{Overview: ASR and spelling correction.} 
\label{fig:framework}
\end{figure}

To enhance the user experience with accurate intent transcription, it is fundamental to enhance the performance of the ASR model \cite{kaldi_optimization}. Typically, the ASR model first generates multiple transcription hypotheses from the audio, then selects the hypothesis with the fewest errors using a language model (LM). Finally, spelling correction is applied to refine the generated sentence and lower the word error rate (WER). Considering the input and output of ASR correction are both texts, utilizing the sequence-to-sequence method for spelling correction ~\cite{10.1007/978-3-642-39593-2_7,d2016automatic,mani2020asr,liao2020improving} becomes a popular direction. 
On the other hand, the word-level or character-level approaches concentrate on distinguishing whether a word or character is meaningful for its surrounding context ~\cite{zhang2020spelling}, providing a context-aware solution for effective correction.
In this paper, we concentrate on \textbf{improving the character-level approaches using contextualized language models}.

The application of deep neural language models~\cite{peters2018deep,radford2019language,devlin2018bert} has gained great success in recent years. 
They create contextualized token representations that are sensitive to the surrounding context. 
The success of contextualized token representations~\cite{,hofstatter2019enriching,10.1145/3397271.3401044,sahrawat2020keyphrase,taille2020contextualized} suggests that despite being trained with only a language modelling task, they learn highly transferable and task-agnostic properties of language. 
With the development of pretraining methods, several attempts~\cite{duan2011online,hagen2017large} have tried to adopt pre-trained language models (PLMs) in Speech Search Query Correction.
\citet{zhang2020spelling} proposed Soft-Masked BERT, which is composed of a detection network and a correction network based on BERT, to solve the Chinese Spelling Correction task.
More recently, \citet{leng2021fastcorrect} introduced FastCorrect-2, an error correction model that leverages multiple candidate inputs to enhance correction accuracy. 

\begin{figure}[!t]
\centering
\includegraphics[scale=0.55]{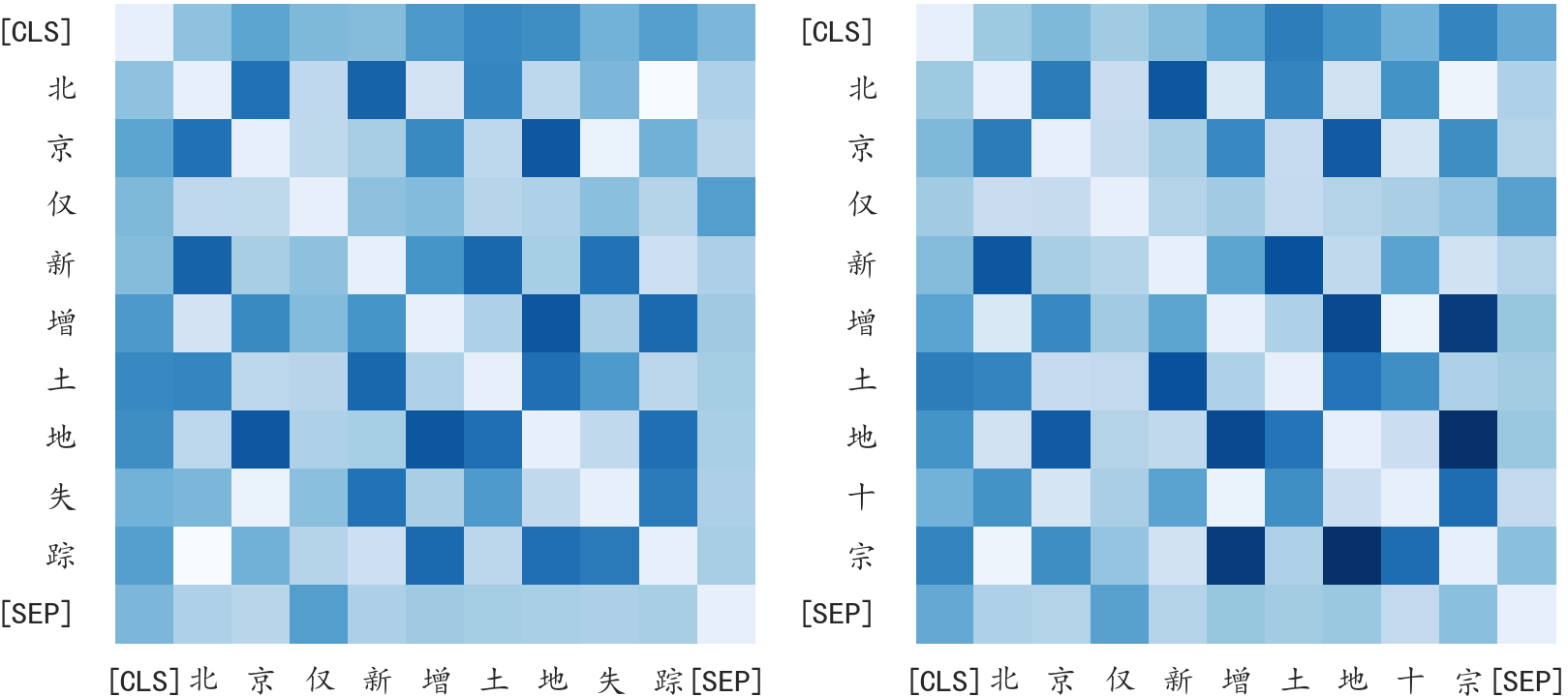}
\caption{The left picture shows the cosine similarity of tokens in a misrecognized query, while the right side describes the reasonable query. A darker color indicates a higher correlation. It makes sense that ``\begin{CJK}{UTF8}{gbsn}新增土地\end{CJK}'' (new land) is more correlated with ``\begin{CJK}{UTF8}{gbsn}十宗\end{CJK}'' (ten) rather than ``\begin{CJK}{UTF8}{gbsn}失踪\end{CJK}'' (missing) after contextualized encoder.}
\label{fig:sim}
\end{figure}

However, these existing models often fail to fully account for contextual information inherent in speech data \cite{pundak2018deep, hong2025dialin}, leaving room for further improvements. As shown in Figure \ref{fig:sim}, we illustrate the cosine similarity differences between each token in a speech query. By extracting the contextualized token representations from the final layer of a fine-tuned BERT model, we observe that correct tokens exhibit stronger correlations with their surrounding context, whereas erroneous tokens tend to misalign with other tokens, highlighting the potential for context-aware improvements in error correction.

This work introduces \textbf{Contextualized Token Discrimination (CTD)}, a BERT-based method for correcting misrecognized query words in ASR systems. Unlike context-free methods~\cite{yu2014chinese,yu2014overview,xiong2015hanspeller,wang2018hybrid}, our approach utilizes enhanced contextualized representations to correct spelling errors based on semantic information. Based on the observation in \cite{ethayarajh2019contextual} that upper layers of contextualized models produce context-specific representations, which can be affected when a query word mismatches its surrounding context, we propose a composition layer, aggregating the input token, contextualized representation, and difference vector to enhance the correction.

The main contributions of this paper include: (1) We introduce a BERT-based method that corrects misrecognized query words using enhanced contextualized representations. (2) We propose a novel composition layer that aggregates input tokens, contextualized representations, and difference vectors, improving correction through a Multi-Layer Perceptron (MLP). (3) We benchmark our approach against various baselines using the SIGHAN Chinese Spelling Error Correction dataset, showing significantly superior performance. (4) We introduce a new benchmark dataset with annotations from realistic audio sources, evaluating our method on thousands of queries with misrecognized tokens, demonstrating notable improvements in correcting grammar and logical inconsistency, and providing practical advantages for industrial deployment.

\section{Related Work}
\label{sec:related}

\subsection{End-to-End ASR System}
End-to-end ASR models have gained increasing popularity in recent years as a way to fold separate components of a conventional pipeline ASR system (i.e., acoustic, pronunciation, and language models) into a single neural network~\cite{synnaeve2019end, wei2024acoustic}. 
RNN transducer (RNN-T)~\cite{rnn-transducer-2012,rnn-transducer-2018} is one of the promising end-to-end ASR models and has drawn more attention recently. 
RNN-T consists of three major components: an encoder network, a prediction network, and a joint network.
The encoder network maps input acoustic frames, i.e., filter-banks or raw waveforms, into a higher-level representation, and the prediction network generates tokens conditioned on the history of previous predictions.
Then the joint network merges the latent representations from the encoder network and the prediction network, and finally decodes the proper tokens.

Recently, Recurrent Neural Network (RNN), Convolutional Neural Network (CNN), and Transformer based on self-attention~\cite{vaswani2017attention} have enjoyed widespread adoption for modeling sequences across various applications \cite{pmlr-v260-hong25a, kwon2017audio}. 
Each of the three architectures has limitations. RNNs are less effective and efficient than Transformers in modeling long dependencies. CNNs exploit local information and are the standard computational block in vision. \citet{schneider2019wav2vec} introduces a CNN that processes raw audio and optimizes it via a next-time-step prediction task. However, local connectivity requires many more layers or parameters to capture global information. Conversely, Transformers excel at modeling long-range global context but are less effective at extracting fine-grained local feature patterns.
Since the milestone breakthrough of masked language modeling~\cite{devlin2018bert}, \citet{baevski2020wav2vec} presented a Transformer~\cite{vaswani2017attention} based framework, Wav2vec 2.0, for self-supervised learning from raw audio by contrastive learning.
\citet{zhang2020pushing} further change the Transformer encoder to a more advanced Conformer encoder~\cite{gulati2020conformer} and propose a recipe for pretraining.

\subsection{Spelling Error Correction}

As is shown in Figure \ref{fig:framework}, spelling error correction is used as a post-processing method to improve the quality of recognized text~\cite{tanaka2018neural,anantaram2018repairing,wu2022phonetic}. 
The word-level or character-level approaches concentrate on distinguishing whether a word or character is meaningful in its surrounding context.
\citet{zhang2020spelling} propose a novel neural architecture, SoftMasked BERT, to detect the incorrect characters in a sequence and replace them with the correct characters.
On the other hand, considering that the input and output of ASR correction are both texts, the sequence-to-sequence method becomes a popular direction. 
\citet{10.1007/978-3-642-39593-2_7} leverage statistic machine translation and \citet{d2016automatic} use phrase-based machine translation system for ASR correction. 
With the development of the attention mechanism, \citet{mani2020asr} trains an autoregressive correction model with a Transformer architecture, and \citet{liao2020improving} incorporates the pre-training method into ASR correction. 

Existing correction methods typically correct errors in a single sentence, relying on its own context. \citet{gong2023enhancing} introduces POS-ARAN, an adjacent relation attention network for question classification, enhancing context representations with POS information and neighboring signals. \citet{li2022past} proposes an error-driven contrastive probability optimization approach for Chinese spell checking. Despite their utility, detecting and correcting errors in natural language, particularly Chinese, remains an unresolved challenge.
Recent years have witnessed a paradigm shift in spelling error correction, evolving from traditional approaches to leveraging pre-trained language models (PLMs) and large language models (LLMs) \cite{wu2023rethinking, tang2023pre, liang2023disentangled, wei2024asr, wei2024acoustic}. These models leverage their contextual understanding and extensive training data to detect and correct spelling errors more effectively \cite{li2024rethinking, wang2024lm, 10.1007/978-3-030-73200-4_36}, with research particularly focused on optimizing and evaluating models' spelling correction to enhance ASR quality through direct utility (or Model-as-a-Service paradigm) \cite{asano2025contextual, udagawa2024robust} or LLM-in-the-loop integration \cite{hong2025llm, chen2023hyporadise}.

\begin{figure*}[!t]
\centering
\includegraphics[scale=0.44]{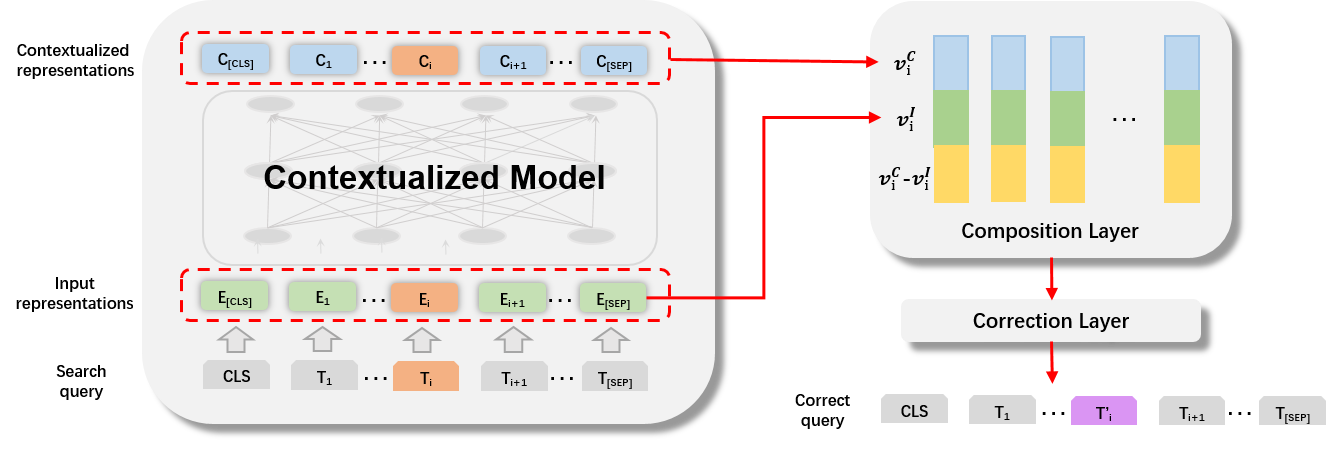}
\caption{Our proposed Contextualized Token Discrimination (CTD) method consists of two main components: (1) a contextualized encoder that produces token representations, and (2) a correction module that enhances these representations using the first and last encoder layers and predicts the correct query. The orange colored token $T_N$ represents an error token that mismatches its context, while the purple $T'_N$ denotes a contextually appropriate token.}
\label{fig:model}
\end{figure*}

\section{METHODOLOGY}
\label{sec:model}

In this section, we introduce the proposed Contextualized Token Discrimination (CTD) method with a pretrained contextualized encoder and a correction module in Speech Search Query Correction.

\subsection{Pretrained Contextualized Encoder}
Since the milestone of \citet{mikolov2013efficient,mikolov2013distributed} was published, a new era in NLP started on word embeddings, also referred to as word vectors.
Word embeddings can learn many properties of a word from large corpora and become one of the most promising methods to capture semantic information in NLP tasks.
More recently, as exemplified by BERT~\cite{devlin2018bert}, ELMo~\cite{peters2018deep}, and GPT-2~\cite{radford2019language}, incorporating context into word embeddings has proven to be a groundbreaking advancement in NLP. Contextualized word embeddings aim at capturing word semantics in different contexts to address the issue of polysemy and the context-dependent nature of words.
Detecting token errors in a search query typically requires a thorough understanding of the surrounding context to ensure both logical coherence and fluency \cite{10.1145/2452376.2452420}. Thus, utilizing pre-trained contextualized representations provides a straightforward solution to the query correction.

In this paper, our contextualized model follows the BERT architecture.
BERT~\cite{devlin2018bert} is a kind of Masked Language Model (MLM), which masks 15\% of words in a sentence, inputs the masked sequence of tokens, and uses the Transformer~\cite{vaswani2017attention} encoder to learn how to use information from the entire sentence to deduce what words are missing.
The main procedures of our pre-trained contextualized representation can be divided into two steps.
Firstly, we fine-tune BERT with MLM loss on correct downstream sentences to learn the correct word dependency.
Then, in the Speech Search Query Correction task, we take the entire search query as input and apply BERT to generate a contextualized representation for each token.
As shown in Figure \ref{fig:model}, we follow the general input style in BERT, which adds ``[CLS]'' and ``[SEP]'' tokens at the beginning and the end of the input sequence, respectively.

\begin{table*}[ht]
\makeatletter\def\@captype{table}
\centering
\scalebox{0.8}{
\begin{tabular}{@{}cccccccccccc@{}}
\toprule
\multirow{2}{*}{Model} & \multicolumn{4}{c}{SIGHAN}   & \multicolumn{4}{c}{AAM}
\\ \cmidrule(l){2-5} \cmidrule(l){6-9}
                  & \multicolumn{1}{c}{Acc.(\%)} 
                  & \multicolumn{1}{c}{Prec.(\%)} 
                  & \multicolumn{1}{c}{Rec.(\%)} 
                  & \multicolumn{1}{c}{F1.(\%)} 
                  & \multicolumn{1}{c}{Acc.(\%)} 
                  & \multicolumn{1}{c}{Prec.(\%)} 
                  & \multicolumn{1}{c}{Rec.(\%)} 
                  & \multicolumn{1}{c}{F1.(\%)} 
                  \\ \midrule
BERT & 76.6 & 65.9 & 64.0 & 64.9 & 40.8 & 35.2 & 31.5 & 33.2 \\
Soft-Masked BERT & 77.4 & 66.7 & 66.2 & 66.4 & 41.2 & 35.7 & 32.0 & 33.7 \\
\midrule
Base Model (Ours) & 93.2 & 92.1 & 86.6 & 89.3 & 55.5 & 53.2 & 48.6 & 50.7 \\
CTD (Ours) & 93.5 & 92.8 & 86.7 & 89.6 & 57.8 & 55.0 & 49.8 & 52.2 \\
\bottomrule
\end{tabular}}
\caption{Experiment results on SIGHAN and AAM datasets.}
\label{tab:res}
\end{table*}

\subsection{Correction Module}
An intuitive way to discriminate troublesome tokens is to distinguish the semantic meaning between the token itself and the token in the context.
Inspired by the aggregated method in Enhanced Sequential Inference Model (ESIM)~\cite{chen2017esim}, we explore a composition layer to construct the token-level discrimination vector between the input representation and the contextualized representation in order to determine whether the input representation and the contextualized token representation are compatible for the current context.

Our composition layer is shown in Figure \ref{fig:model}. 
We concatenate the source token representation, the contextualized representation, and the difference vector as follows:
\begin{equation}
    c_i = [v^C_i;v^I_i;v^C_i - v^I_i],
\end{equation}
where a heuristic matching approach~\cite{chen2017esim} with difference is used here to obtain aggregated information $c_i$ at the $i$-th token position. 
$v^I_i$ is the $i$-th input token representation, while $v^C_i$ is the $i$-th contextualized representation.
We obtain $v^C_i$ from the last layer of BERT.

After aggregating token-level composition information $c_i$, we conduct a classifier to distinguish the proper token according to the aggregated token representations.
For each token $T_i$, the probability of error correction is defined as:
\begin{equation}
    p(y_i|T_i) = softmax(Wc_i + b),
\end{equation}
where $p(y_i|T_i) \in \mathbb{R}^V$ indicates the conditional probability;
$W \in \mathbb{R}^{V \times 3d}$ is weight matrix;
$b \in \mathbb{R}^V$ is bias;
$V$ is the vocabulary size, and $d$ indicates the dimension of BERT.

In our CTD method, end-to-end training is conducted, and we only compute the loss on specific tokens for training.
Firstly, we obtain error tokens $\Lambda_1$ by collecting the differences between the error query and the ground-truth query, formulated as $\forall T_i \ne T'_i \in \Lambda$.
Secondly, we randomly sample extra tokens $\Lambda_2$ for training in order to provide the information on which tokens are correct in the original sequence.
We format $\Lambda_2$ as $\exists T_i = T'_i \in \Lambda$.
In general, $\Lambda = \Lambda_1 + \Lambda_2$ and $\Lambda_1 : \Lambda_2 = 1:5$. The training objective is thus formulated as:
\begin{equation}
    L = - \sum_{n=1}^{N}\sum_{i \in \Lambda} \log p(y_i|T_i),
\end{equation}
where N is the number of training examples.

\section{Experiments and Results}
\label{sec:Experiment}

In this section, we conduct experiments on two Chinese datasets: (1) the publicly available benchmark dataset \textbf{SIGHAN}\footnote{\url{http://ir.itc.ntnu.edu.tw/lre/sighan8csc.html}} and (2) our newly built \textbf{AAM}\footnote{The full dataset will be publicly released upon acceptance.}. Both datasets consist solely of text pairs: speech query with potential homophonic errors is paired with the corresponding correct sentence. In all experiments, we input the query into our model and treat the correct sentence as the ground truth.
The pretrained contextualized representations are refined during model training. For evaluation, we measure the sentence-level accuracy, precision, recall, and F1 score \cite{udagawa2024robust}.

\subsection{SIGHAN Task}
\label{ssec:sighan}

SIGHAN is a public dataset containing 1,100 texts and 461 types of errors (characters), widely used in spelling error correction tasks \cite{tseng2015introduction}.
We adopted the standard split of training, development, and test data of SIGHAN.

Since BERT and Soft-Masked BERT~\cite{zhang2020spelling} are two primary methods we are compared to, we align to previous experimental settings and results. 
For our base model, we fine-tune the BERT-base-chinese model\footnote{\url{https://huggingface.co/bert-base-chinese}}, which consists of 12 self-attention layers, a hidden state dimension of 768, and 12 attention heads per layer.
Moreover, we employ a linear classifier at the top layer to predict a proper token at each position.
The linear layer has a dimension of 768, and the dictionary size is set to 21128.
With this configuration, our fine-tuned base model has 120M parameters in total. Furthermore, we adopt a composition layer to aggregate token information and use a linear classifier with 2304 units.
In this setting, our contextualized token discriminator (CTD) has 156M parameters. 

For model training, we use AdamW optimizer~\cite{loshchilov2017decoupled} with $\beta_1 = 0.9$, $\beta_2 = 0.98$ and $\epsilon= 10^{-9}$.
Transformer learning rate schedule~\cite{vaswani2017attention}, and the learning rate is set to rise linearly from 1e-7 to 5e-4 in the first 5\% steps, then exponentially decay it in the rest of the training procedure~\cite{goyal2017accurate}.
All of our experimental models are trained with 512 sequences per mini-batch, and the early-stopping strategy is adopted in all the experiments.

As shown in Table~\ref{tab:res}, our base model achieves an F1 score of 89.3\% on the SIGHAN test set. We observe significant improvements over the previous BERT and Soft-Masked BERT benchmarks, which is surprising given that we only fine-tuned the MLM using the training set and adjusted the hyperparameters. The domain-specific data likely helps BERT distinguish correct tokens through careful fine-tuning of the MLM. Additionally, the F1 score increases to 89.6\% after incorporating our proposed CTD. Based on this case study, we conclude that performance is limited by the availability of domain-specific data, motivating a larger and more diverse benchmark to validate the generalizability of our method and contribute meaningfully to the evaluation of future speech query correction research \cite{wei2024asr, hong2025qualbench}.

\begin{CJK}{UTF8}{gbsn}

\begin{table*}
    \centering
    \small
    \renewcommand\arraystretch{1.5}
    \begin{tabular}{cc} 
    \toprule
    Input & 
    \begin{CJK}{UTF8}{gbsn}
    楼市调控的行政手段\textbf{\underline{意见}}不宜加。
    \end{CJK} \\
    & (Administrative measures and \textbf{\underline{opinions}} on property market regulation should not be increased.) \\
    \hline
    Ours & 
    \begin{CJK}{UTF8}{gbsn}
    楼市调控的行政手段\textbf{\underline{宜减}}不宜加。
    \end{CJK} \\
    & (Administrative measures for property market regulation \textbf{\underline{should be reduced}} rather than increased.) \\
    \bottomrule \toprule
    Input & 
    \begin{CJK}{UTF8}{gbsn}
    北京仅新增住宅土地供应\textbf{\underline{失踪}}。
    \end{CJK} \\ 
    & (Only new residential land supply \textbf{\underline{missing}} in Beijing.) \\
    \hline
    Ours & 
    \begin{CJK}{UTF8}{gbsn}
    北京仅新增住宅土地供应\textbf{\underline{十宗}}。
    \end{CJK} \\
    & (Only \textbf{\underline{ten}} new residential land supply in Beijing.) \\
    \bottomrule
    \end{tabular}
    \caption{Examples from the AAM test set. ``Input'' refers to a sentence that may contain errors, while ``Ours'' indicates the top-1 correction result produced by our proposed CTD method.}
    \label{tab:examples}
\end{table*}
\end{CJK}

\begin{table}[!t]
\makeatletter\def\@captype{table}
\centering
\scalebox{0.8}{
\begin{tabular}{cccc}
\hline
Datasets  & \makecell{Duration\\(Hours)} & Evaluation CER & Generated Pairs \\ \hline
Aidatatang & 140              & 8.23           & 823                \\ 
Aishell-1  & 151              & 6.75           & 1475               \\ 
Magicdata  & 712              & 5.58           & 4657               \\ \hline
Total     & 1003             & 6.85           & 6965               \\ \hline
\end{tabular}}
\caption{Details of the AAM dataset.} 
\label{tab:aam}
\end{table}

\subsection{AAM Task}
In this task, we collect misrecognized cases from ASR systems in order to further verify the effectiveness of our proposed method in more realistic cases.
We carry out our experiments by annotating public speech resources: \textbf{A}idatatang\footnote{\url{http://www.openslr.org/62/}}, \textbf{A}ISHELL-1\footnote{\url{https://www.openslr.org/33/}} \cite{bu2017aishell} and \textbf{M}agicData\footnote{\href{http://www.imagicdatatech.com/index.php/home/dataopensource/data_info/id/101}{www.imagicdatatech.com/home/dataopensource}}.
We denote this task as \textbf{AAM}.

Since the three datasets contain substantial audio and transcripts commonly used in Mandarin speech recognition tasks, we employ a practical ASR model to generate erroneous hypotheses paired with ground-truth text. Inspired by \cite{gulati2020conformer}, we use the Convolution-augmented Transformer (Conformer) \cite{gulati2020conformer} as our acoustic model. We train a baseline Conformer\footnote{\url{https://github.com/hirofumi0810/neural_sp}} on each corpus individually. As shown in Table \ref{tab:aam}, we successfully train three Conformer models with low character error rate (CER) using the ESPnet toolkit~\cite{li2020espnet}, allowing us to generate audio transcriptions and extract the erroneous cases for constructing the evaluation dataset.

\paragraph{Dataset overview.} Overall, the AAM dataset consists of 6965 pairs (4965 training data, 1000 validation data, and 1000 test data).
It is also noted that we only keep pairs of erroneous hypotheses and ground-truths of the same length to follow the style of the SIGHAN task.
We do not consider the errors of insertions and deletions.
Moreover, we filter erroneous hypotheses with misrecognized stopwords because stopwords have almost no influence on the contextualized representation.

\begin{figure}[!t]
\centering
\includegraphics[scale=0.25]{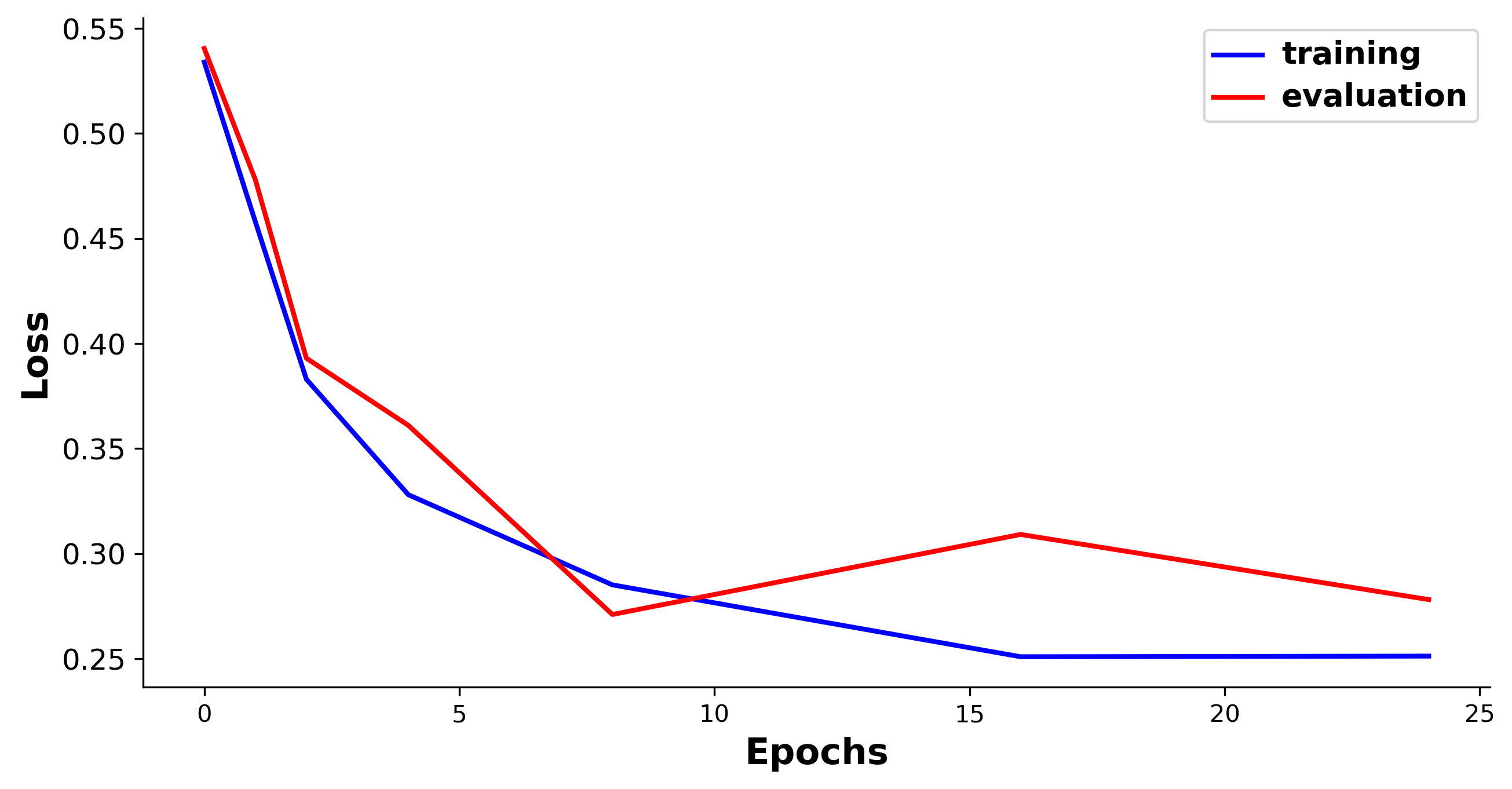}
\caption{Loss of our CTD on the AAM data}
\label{fig:corrector_loss}
\end{figure}

\paragraph{Results.} We reuse the BERT checkpoint, model architecture, and hyperparameters as described in Section~\ref{ssec:sighan}. Our CTD model is trained for 25 epochs, and the results are illustrated in Figure~\ref{fig:corrector_loss}. As shown in Table~\ref{tab:res}, our well-trained methods (i.e., base model and CTD) achieve comparable results to both the previous BERT benchmark and Soft-Masked BERT, with F1 scores of 50.7\% and 52.2\%, respectively.

To demonstrate the effectiveness of our CTD method, we present several cases in Table~\ref{tab:examples} from the AAM test sets.
For example, there is a typo in the sentence
``\begin{CJK}{UTF8}{gbsn}楼市调控的行政手段意见不宜加\end{CJK}''. The noun ``\begin{CJK}{UTF8}{gbsn}意见\end{CJK}'' (opinions) in the context is illogical and should be written as a word ``\begin{CJK}{UTF8}{gbsn}宜减\end{CJK}'' (should be reduced). ``\begin{CJK}{UTF8}{gbsn}意见\end{CJK}'' and ``\begin{CJK}{UTF8}{gbsn}宜减\end{CJK}'' are easily misrecognized by ASR models due to their similar pronunciation in Chinese. In the second case, the phrase "land supply missing" makes no sense. We find that our CTD method can correct such typos by referencing grammar and logical consistency. However, it remains very challenging for errors that require world knowledge, which current AI systems struggle to handle effectively.

\section{Conclusion}
\label{sec:Conclusion}
In this paper, we propose a simple and effective method, Contextualized Token Discrimination (CTD), to enhance model performance for speech search query correction. We collect thousands of erroneous queries from different speech recognition datasets for comprehensive benchmarking. The experiments show the superior performance of our method over previous works. In the future, we aim to explore better subcultural approaches to addressing semantic errors and visualizing the differences between correct and incorrect representations.

\section*{Limitations}

While our proposed method demonstrates significant advancements in speech search query correction, and the proposed dataset captures a realistic evaluation of error correction with sufficient complexity, this work has two main limitations. First, the AAM dataset focuses solely on same-length erroneous and ground-truth pairs, limiting its ability to handle insertion and deletion errors, which are also common in ASR transcriptions and can significantly impact the accuracy of speech recognition systems. Second, the proposed method’s reliance on domain-specific training data may constrain generalization to diverse contexts, motivating the design of few-shot learning approaches to further reduce deployment costs.

\bibliography{custom}

\end{document}